\pdfoutput=1

\documentclass[12pt]{article}

\usepackage{graphicx,amssymb,amsmath,subfig,amsfonts}
\usepackage{url}
\usepackage{natbib}
\usepackage{hyperref}
\usepackage{cleveref}
\usepackage{caption}
\newcommand{\e}{\text{e}}
\newcommand{\mtx}[1]{\mathbf{ #1}}
\newcommand{\E}[1]{{\mathrm E}\left[ #1 \right]}

\newcommand{\diag}[1]{\mathrm{diag}\left( #1\right)}

\title{\vspace{-15mm}%
	\fontsize{24pt}{10pt}\selectfont
	\textbf{Methods and Concepts in Economic Complexity}
	}	
\author{%
	\large
	\textsc{Andres Gomez-Lievano} \\[1.5mm]
	\normalsize	Growth Lab, Center for International Development \\
	\normalsize	Harvard University 
	\vspace{-5mm}
	}
\date{}

\begin{document}

\maketitle

\begin{abstract}
Knowhow in societies accumulates as it gets transmitted from group to group, and from generation to generation. 
However, we lack of a unified quantitative formalism that takes into account the structured process for how this accumulation occurs, and this has precluded the development of a unified view of human development in the past and in the present. 
Here, we summarize a paradigm to understand and model this process. 
The paradigm goes under the general name of the Theory of Economic Complexity (TEC). 
Based on it, we present a combination of analytical, numerical and empirical results that illustrate how to characterize the process of development, providing measurable quantities that can be used to predict future developments. 
The emphasis is the quantification of the collective knowhow an economy has accumulated, and what are the directions in which it is likely to expand. 
As a case study we consider data on trade, which provides consistent data on the technological diversification of 200 countries across more than 50 years. 
The paradigm represented by TEC should be relevant for anthropologists, sociologists, and economists interested in the role of collective knowhow as the main determinant of the success and welfare of a society. 
\end{abstract}

\newpage
\tableofcontents

\newpage

\section{Introduction}
The Theory of Economic Complexity (TEC) promises to become a competing alternative to understand the evolution and development of human societies. TEC is the basis of the Scrabble Theory of Economic Development (STED). While the TEC paradigm has inspired a lot of work in the last decade addressing important questions in economic development and economic geography, here we limit our discussion to the general mathematical and methodological formalisms implied by it.

The idea underpinning TEC (and STED) is that socioeconomic development is the result of accumulating, coordinating and deploying increasing amounts of knowhow in a society \citep{HidalgoEtAl2007,HidalgoHausmann2009,HausmannHidalgo2011}. This idea is rooted in three characteristics about human biology: humans are limited in their capacity to reason (``bounded rationality''), in their capacity to learn (this ``bounded learnability'' partly inducing specialization), and in their capacity to transmit knowledge (giving rise to the notion of ``tacit knowledge''). These limitations are the main forces that drive people to come together in teams of individuals in order to \emph{combine} their individual tacit knowhow.\footnote{Coming together to collaborate with one another is in contrast coming together to learn from one another. The latter, as we will see, plays a lesser role in economic development, in spite of the fact that learning certainly happens and is an externality that economists have studied extensively.} The process of accumulating and coordinating capabilities into increasingly complex productive endeavors has driven cultural evolution of societies in the past \cite{henrich2015secret}, but is of relevance since still today is what drives the economic development of nations, regions and cities. 

TEC assumes that complex production processes are those that combine a large multiplicity of different, but complementary, capabilities. Places (e.g., cities) that are productive are those in which capabilities are abundant and can combine with relative ease \citep{gomez2016explaining,neffke2017coworker}. In brief, economic development is a process of collective learning (in contrast to, or in addition to, a collective process of individual learning).

The assumption of bounded learnability has a number of implications for how technology moves and concentrates in space. While it is hard for individuals to learn, collectives (firms, cities, countries) on the other hand can learn by attracting people. Hence, the diffusion of technology does not move necessarily because of ideas flowing into people heads, but because of people with ideas migrating. Capabilities embodied in the brain of people will move where they can combine with other capabilities in a productive manner. Precisely because ideas do not ``spillover'' easily from person to person, there will be a disproportionate accumulation of capabilities in few places. The theory thus suggests why technologies accumulate slowly, and why they are unequally distributed in space. 

In the next section we create a simple mathematical model of economic complexity that codifies these ideas mathematically. The third section is devoted to developing some frameworks to think about the implications for the movement of capabilities and the diffusion of technology.


\section{Simple model of economic complexity}\label{sec_model}
Let places (e.g., cities) be indexed by $c$, economic activities be indexed by $p$ (e.g., industry specific output or product $p$), and firms by $i$. Given these (cities, products, and firms) let us try to derive an expression for the probability that a firm $i$ is able to produce product $p$ in city $c$: $$ \Pr(X_{i,c,p}=1)~=~~?,$$
where $X_{i,c,p}$ is simply a variable that takes the value of 1 if the firm $i$ is able to operate successfully in the city, and 0 if not.

We will assume that an entrepreneur will be able to run a firm $i$, that is, she will be able to operate in city $c$ and produce the industry-specific product $p$, depending on whether she is able to put together the ``right'' knowhow. Assume this implies combining $M_p$ different and complementary capabilities. We will leave these capabilities unspecified. The important part here is that capabilities are all the ingredients needed to produce the product $p$. These include, in principle, knowhow of finance, legal issues, engineering, research and development, sales and marketing. Thus, we will typically think of ``capabilities'' as ``professional or job occupations'', although they can also include public services that a production process may need as a necessary requirement.

The parameter $M_p$ represents, in this view, the ``complexity'' of the economic activity associated with the production of $p$. The more capabilities are needed, the larger the value of $M_p$, and the more complex the activity. Notice that this approach differs from the conventional production process assumed in economics, whereby the emphasis is on the substitutability of a few production factors (e.g., capital and labor). Instead, we are assuming that (i) there is no substitutability between capabilities, and that (ii) the number of factors is not two, but $M_p \gg 1$. The reason we need to think probabilistically in this model comes from this assumption about the large multiplicity of capabilities \citep{gomez2016explaining}. 

Let $s_i$ be the probability that the entrepreneur of firm $i$ has any random capability of the $M_p$ capabilities required by the business.\footnote{In other words, the number of capabilities the entrepreneur is expected to have is, on average, $s_i M_p$.} This probability can be interpreted as a measure of the entrepreneur's individual knowhow. For example, she may be trying to open a firm that will manufacture $p=\text{shoes}$, which let us assume requires $M_p=10$ different capabilities, and $s_i$ represents the fact that she has the capacity to easily act both as a designer and a manager, so $s_i\approx 2/10=0.2$. The larger the parameter $s_i$ is, the better equipped is the entrepreneur in operating the business individually, the less she needs a team of people supporting her as a consequence, and the less dependent she will be of the city she lives in and on what the city offers to her. Notice, however, that while $s_i$ can be interpreted as ``the level of schooling'', it does not track the \emph{depth} of knowledge but the \emph{breadth}: It is about how many different things she could know how to do individually. The probability she will be able to operate successfully the firm, all on her own, is $s_i^{M_p}$. Since $s_i$ is a number between 0 and 1, the more complex the economic activity, the probability she will be successful will decrease exponentially. But how does the city change the probability of the entrepreneur to be able to run her business, $\Pr(X_{i,c,p}=1)$?

Of the $M_p$ capabilities required to produce product $p$, suppose the city $c$ ``provides'' $D_c$ capabilities to the entrepreneur $i$ (where $0\leq D_c\leq M_p$). In other words, through the family, friends, colleagues and, in general, public and private services, which she is typically exposed to on a regular basis by living in city $c$, the entrepreneur could in principle be able to get and complete the remaining $8/10$'s missing skills and capabilities outside her expertise, which she expects to require to run her firm. Presumably, the bigger the city, the more diverse, and the larger $D_c$ will be, and the easier it will be to get those capabilities. 

The problem, note, is that the entrepreneur will only get \emph{all} the $M_p$ capabilities she needs if the $D_c$ capabilities offered by the city \emph{contain} the capabilities that she does not have given $s_i$. The only situation in which the entrepreneur will be able to run her business is if she requires \emph{none} of the capabilities the city \emph{does not} have. 

Let us say this again. By living in city $c$, the entrepreneur can be sure she has $D_c$ of the $M_p$ capabilities. These are a given, in a sense, and she does not need to worry (too much!) about them. The challenge she faces is rather with getting the $M_p - D_c$ capabilities not offered by the city, which she cannot take for granted. These are capabilities that she will need to bring to the business on her own, without the help of the city. Necessarily, solving the challenge of lacking $M_p - D_c$ capabilities will depend on her own individual knowhow. She has a probability $s_i$ of having any of those capabilities. 

Let us compute the probability that she will be able to get them all and operate her firm 
$$\Pr(\text{firm $i$ in $c$ produces $p$}~|~\text{city facilitates $D_c$ of the $M_p$ capabilities}),$$
which we can write more concisely as $\Pr(X_{i,c,p}=1~|~D_c)$. According to the reasoning above, this probability is equal to
\begin{align}
	\Pr(X_{i,c,p}=1~|~D_c)&= s_i^{M_p-D_c}.
	\label{eq_probgivenDc}
\end{align}
\Cref{eq_probgivenDc} is the product $s_i\times s_i \times \cdots \times s_i $ because it is the probability of having the first capability times the probability of having the second, and so on, until we have the probability of getting \emph{each} of the missing capabilities not offered by the city.

In reality, however, $D_c$ is not a fixed number. Cities are messy places, they change from neighborhood to neighborhood and from day to day, and no person knows the city as a whole completely. Hence, if our entrepreneur is very unlucky she may get $D_c=0$, or she can be super lucky and get $D_c=M_p$. To take this stochasticity into account, we can think instead of the \emph{probability} that the city provides \emph{any} of the capabilities. Let us denote this probability by $r_c$. The expected number of capabilities required to produce $p$ that the city can offer on average is $\E{D_c} = r_c M_p$. Thinking of $D_c$ probabilistically, means thinking of $D_c$ in this context as a ``binomially distributed random variable'' with parameters $M_p$ and $r_c$. 

To correctly compute the probability that our entrepreneur will be able to manage her business we need to average \Cref{eq_probgivenDc} over all the possible number of capabilities the city may offer:
\begin{align}
	\Pr(X_{i,c,p}=1) &= \sum_{D=0}^{M_p}\Pr(X_{i,c,p}=1~|~D)\Pr(D) \nonumber \\
	&= \sum_{D=0}^{M_p}s_i^{M_p-D}\binom{M_p}{D}r_c^{D}(1-r_c)^{M_p-D} \nonumber \\
	&= \sum_{D=0}^{M_p}\binom{M_p}{D}r_c^{D}(s_i(1-r_c))^{M_p-D} \nonumber \\
	&= (r_c + s_i(1-r_c))^{M_p}.
	\label{eq_prob}
\end{align}
For clarity, let us denote this probability as an explicit function of the parameters involved, $\Pr(X_{i,c,p}=1)\equiv f(M_p, s_i, r_c)$. Using the properties of exponentials and logarithms, \Cref{eq_prob} can be simplified to yield the following expression:
\begin{align}
	f(M_p, s_i, r_c) &\approx \e^{ - M_p (1-s_i)(1-r_c)}.
	\label{eq_approx}
\end{align}

\subsection{Insights}
\Cref{eq_approx} contains several insights. First, let us recall the meaning of the terms again:
\begin{enumerate}
	\item $M_p$: This is the number of capabilities required to produce the industry-specific product $p$. We can refer to it as the ``complexity'' of the product.
	\item $1-s_i$: This is the manager-specific probability of lacking any one of the capabilities required in production processes. Conversely, $s_i$ can be referred to as a measure of ``individual knowhow''.
	\item $1-r_c$: This is the city-specific probability of lacking any one of the capabilities required in production processes. Conversely, $r_c$ can be referred to as a measure of ``collective knowhow'' and represents a measure of input availability, which in turn represents a measure of urban diversity.	
\end{enumerate}
Not surprisingly, increasing any of these three terms will \emph{decrease} the probability that firm $i$ will exist. But the crucial observation is that they involve \emph{exponential} changes. That is to say, small changes in any of these three terms can in principle have (exponentially) large effects on the success of firms, specially if the value of those variables is already high. But let's study the partial rates of change separately, in order to compare them:
\begin{description}
	\item[Technological improvement of production process of $p$:]
	\begin{align}
		\frac{\partial f/\partial (-M_p)}{f}&= (1-s_i)(1-r_c), \label{eq_Mp}
	\end{align}
	
	\item[Individual learning for entrepreneur $i$:]
	\begin{align}
		\frac{\partial f/\partial (M_p s_i)}{f}&= ~ (1-r_c), \label{eq_si}
	\end{align}
	
	\item[Collective learning for city $c$:]
	\begin{align}
		\frac{\partial f/\partial (M_p r_c)}{f}&= ~ (1-s_i) \label{eq_rc}.
	\end{align}
\end{description}
The partial derivatives have the term $M_p$ because we want them to reflect ``changes in the number of capabilities''. Hence, $\partial (-M_p)$ represents the reduction of the number of capabilities required to produce $p$, $\partial (M_p s_i)$ represents the increase in the average number of capabilities known by the individual $i$, and $\partial (M_p r_c)$ represents the increase in the average number of capabilities present in city $c$. In other words, the probability the firm will be successful, $f(M_p, s_i, r_c)$, will increase according to \Cref{eq_Mp} if the complexity of $p$ decreases (through technology improvements), \Cref{eq_si} tells us that it will increase if the individual knowhow of the entrepreneur $i$ increases (individual learning), and \Cref{eq_rc} that it will increase if the collective knowhow of the city $c$ increases (collective learning).

\begin{figure}[!th]
	\begin{center}
		\includegraphics[width=0.9\textwidth]{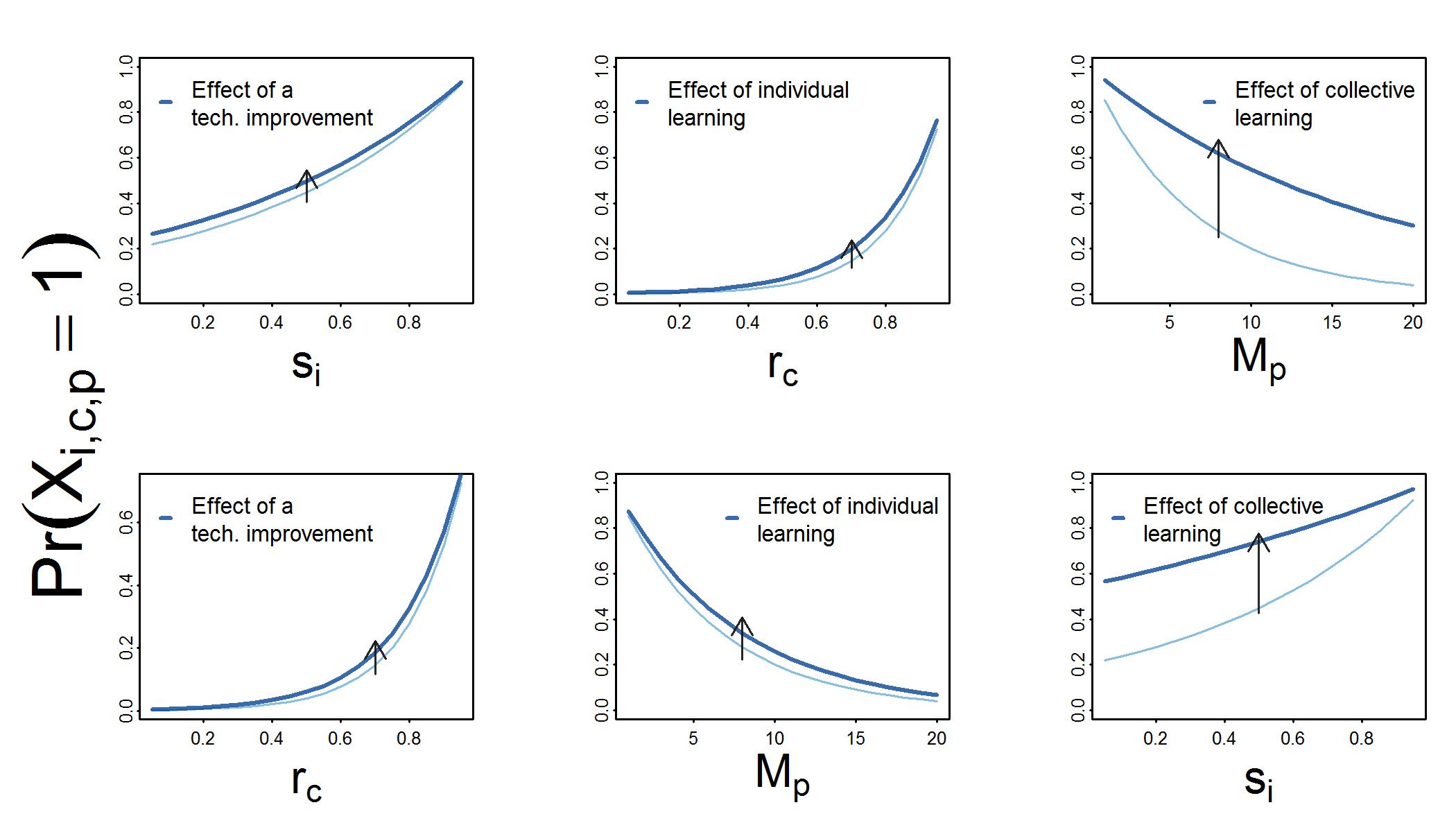}
		\caption{Comparing the different ways of increasing the probability of operating a firm in city $c$ that produces product $p$. The increase in probability represented by the change from lightblue to darkblue is, in each case, due to a change that represents that production is one capability more easy. Hence, a technological improvement is when $M_p$ is reduced by 1, individual learning is when $s_i$ is increased by $1/M_p$, and collective learning is when $r_c$ is increased by $1/M_p$. For each panel, one of the parameters is explicitly shown to vary across the $x$-axis, another parameter is changed in order to represent the change in probability (either by a technological improvement in the left panels, individual learning in the middle panels, and collective learning in the right columns), and another parameter is implicitly kept constant, correspondingly at values $M_p=8$, $s_i=0.2$, or $r_c=0.8$.}
	\label{fig_model_learning}
	\end{center}
\end{figure}

Let us study in detail the magnitude of these rates of change. On the one hand, we have $M_p$ which is supposed to be large, $M_p\gg 1$. On the other hand, $s_i$ and $r_c$ are probabilities and are therefore between 0 and 1. However, since the city is a collective, the probability it provides an input is larger than the probability an individual has it, so $r_c\gg s_i$. Conversely, $1-r_c\ll 1- s_i$. Consequently, we have that $0 < (1-r_c)(1-s_i) < 1-r_c \ll 1 - s_i$. The implication is that these rates have the following order:
\begin{align}
	0<\frac{\partial f/\partial (-M_p)}{f} < \frac{\partial f/\partial (M_p s_i)}{f} \ll \frac{\partial f/\partial (M_p r_c)}{f}
	\label{eq_comparison}
\end{align}
Thus, the effect of a technology improvement is smaller than the effect of individual learning which is much smaller than the effect of collective learning 
$$\text{{\footnotesize Effect of tech. improvement}}<\text{{\footnotesize Effect of individual learning}}\ll\text{{\footnotesize Effect of collective learning}}.$$ Increasing the collective knowhow of a city (e.g., through immigration, direct foreign investments, etc.) has a significant effect on the probability of entrepreneur $i$ being able to operate a business which produces product $p$. \Cref{fig_model_learning} illustrates these effects.

Of course, the comparison in \Cref{eq_comparison} has several problems and hinges on highly simplifying assumptions. For example, the comparison assumes that a linear (infinitesimal) change in the three variables is comparable among them. In other words, it does not take into account the \emph{cost} of these changes. But one can play a bit with the equations, make some assumptions, and it is easy to see that this result holds for a wide range of situations.

A second intuition is that the effects of collective learning is differentially distributed across people and economic activities. The combined effect can be summarized by noticing that cities with a large body of collective knowhow will make ``difficult'' activities easier. And the difficulty can be because the activity is itself very complex, or because the entrepreneur lacks several capabilities, or both. \Cref{eq_rc} shows these two effects clearly given it is a function of both $M_p$ and $s_i$. 

Intuitively, thus, it is easy to see that increases in collective knowhow have a reinforcing effect and suggest a virtuous cycle: a place with a relative large body of collective knowhow will attract more people and facilitate more complex economic activities, which themselves will increase the body of collective knowhow in that place. This process will thus propel a run-away cycle of collective learning that will concentrate economic activities and wealth in a relatively compact region of space: a ``city''. The more complex the activities, the more concentrated they will be across places. This explains why complex innovations tend to happen in large diverse cities \citep{gomez2016explaining}.

\subsection{The distributional implications of multiplicativity versus additivity}\label{sec_multadd}
Our model assumes that a large set of inputs must be combined to generate an output. The output will not be produced, however, if any of the inputs is missing. This is a specific form of a production function called the Leontief production function. By using a Leontief function we are assuming that the presence of an output follows a logic of \emph{complementarity}. Mathematically, complementarity meant taking products of probabilities. Thus, it can also be said that outputs follow a logic of \emph{multiplicativity}. This logic allowed us to calculate the consequences of our model, summarized in \Cref{eq_approx}.

We can claim that the logic of multiplicativity between inputs for determining the presence of an output, in turn, implies that the presence of an input follows a logic of \emph{additivity} between outputs. To see this, let us put both ``logics'' in words. On the one hand, we say that to observe a unit of output of a given product X we need input A \emph{and} input B \emph{and} input C, and so on, such that we list all the inputs required to produce X. On the other hand, this allows us to say that if we observe the presence of input A it is because output X is present \emph{or} output Y is present \emph{or} output Z is present, and so on, as we list all the products that require input A as a necessary factor of production. Thus, concatenating conditions with ``and'''s is akin to multiplying probabilities while concatenating conditions with ``or'''s is akin to adding probabilities. Consequently, outputs follow a logic of multiplicative processes while inputs follow a logic of additive processes. This suggests that magnitudes of production across outputs should be distributed with lognormal-like distributions while the values of presence of inputs should be distributed with normal-like distributions.\footnote{These two logics, of course, are then guided by economic forces of supply and demand.}


\section{The structure and dynamics of collective learning}
We have stated that the process of collective learning is based on the accumulation of productive capabilities. We now turn our attention to \emph{how} places acquire these capabilities. The question that arises is what type of process describes collective learning? 

In the last decade, observations have shown that there is a particular way for how cities, regions, and countries diversify their economic activities. The finding behind these observations consists of the fact that since economic activities are the result of combining capabilities, some economic activities use very similar sets of capabilities. This means that the probability of producing a product $p$ increases if a place also produces products $p'$ which use similar capabilities. A pair of products which use similar capabilities are referred to as ``related'', which is why this particular way of diversification has been recently referred to as the ``principle of relatedness'' \citep{HidalgoEtAl2007,Neffke2013SkillRelatedness,hidalgo2018principle}. 

In addition to the principle of relatedness, there is another phenomenon, analogous to it, which has been found to occur widely: learning typically comes from \emph{imitation}. Imitating others is a successful learning strategy for individuals, for example, when the cost of individually carrying out the research to find the solution to a problem is too costly. That imitation is the basis of learning is foundational in the field of cultural evolution. The importance is not only because it is empirically supported by experiments that show that humans are super-imitators (in contrast to other mammals), but because it is the aspect of human biology that explains the existence of culture. This line of scholarly work has shown that similar cultural traits are acquired by populations that are culturally similar. While this sounds circular and obvious, note that this type of collective learning did not have to occur in this way: a particular society could tend to acquire traits found in the societies that are the most dissimilar to it. Yet this is not what is observed, and the most likely explanation is precisely because learning is costly. The consequence of this is that learning from societies that are ``culturally close'' is a less-risky form of collective learning. We can call this the ``principle of collective imitation''.

The principle of relatedness and the principle of collective imitation can both be used to predict which new products a place will be able to produce in the future, based on a matrix that tells which pairs of products are related, and another matrix which tells us the similarity between places.

Let us express these two principles mathematically: 
\begin{description}
	\item[Principle of relatedness:]
	\begin{align}
		M_{c,p}(t+1) &= \sum_{p'} M_{c,p'}(t)\frac{\phi(p', p)}{\sum_{p''}\phi(p'', p)} \nonumber\\
		&= \sum_{p'} M_{c,p'}(t)P(p', p)\quad\text{where $\sum_{p'} P(p', p) = 1$}.
	\end{align}
	Here, $M_{c,p}(t)$ can be thought of a measure of production of product $p$ by place $c$ at time $t$. The term $\phi(p', p)$ is measure of relatedness between products (e.g., similarity in their production processes), such that it is large if $p'$ and $p$ share several capabilities. We will discuss how to construct this matrix later. For now, let us just remark on the fact that we have assumed that the principle of relatedness implies a weighted average of the products a place $c$ is already producing, weighted by their similarity to $p$. We can write it in matrix form as
\begin{align}
	\mtx{M}(t+1) &= \mtx{M}(t)\cdot \mtx{P},
\label{eq_relatedness}
\end{align}
where the elements of each specific column in $\mtx{P}$ add up to 1. In other words, $\mtx{P}$ is column-normalized. Such matrix is called a ``left-stochastic matrix''.

	\item[Principle of collective imitation:]
	\begin{align}
		M_{c,p}(t+1) &= \sum_{c'} \frac{\chi(c, c')}{\sum_{c''}\chi(c, c'')} M_{c',p}(t) \nonumber\\
		&= \sum_{c'} C(c,c')M_{c',p}(t)\quad\text{where $\sum_{c'} C(c, c') = 1$}.
	\end{align}
	The term $\chi(c, c')$ is measure of similarity between places (e.g., cultural similarity). We will also discuss how to construct this matrix later. Again, let us just remark on the fact that we have assumed that the principle of collective imitation can be modeled as a weighted average of the production of the specific product $p$ across places $c'$, weighted by their similarity to $c$. We can write it in matrix form as
\begin{align}
	\mtx{M}(t+1) &= \mtx{C}\cdot\mtx{M}(t),
\label{eq_imitation}
\end{align}
where the elements of each specific row in $\mtx{C}$ add up to 1. In other words, $\mtx{C}$ is row-normalized. Such matrix is called a ``right-stochastic matrix''.
\end{description}

\subsection{Some comments about stochastic matrices}
Stochastic matrices get their name from the fact that they are the main object one uses to model a wide variety of random processes. A random process is a sequence of random variables, and one typically assumes there are some ``rules'' for how a random variable at a given time-step $t$ changes into a new random variable at time $t+1$. Despite their name, stochastic matrices appear in different instances not necessarily attached to any stochastic process. \Cref{eq_relatedness,eq_imitation} are one such example. Once a stochastic matrix appears, however, it does open the door to thinking about the process in more probabilistic terms. Hence, it is useful for us to make the following two distinctions:
\begin{description}
	\item[Multiplying a \underline{row} vector on the \underline{left} of $\mtx{P}$ (a \emph{left}-stochastic matrix):] 
	Computes weighted averages, $$\vec{x}(t+1)^T = \vec{x}(t)^T\cdot\mtx{P}.$$
	In this specific instance, the elements of the vector $\vec{x}(t)^T$ are usually a property, measure or characteristic that varies across products. This type of dynamics describes a phenomenon in mathematics called ``consensus dynamics''.
	\item[Multiplying a \underline{column} vector on the \underline{right} of $\mtx{P}$ (same matrix as above):] 
	Propagates/diffuses the values of the vector, $$\vec{n}(t+1) = \mtx{P}\cdot\vec{n}(t).$$
	Here, the elements of the vector $\vec{n}(t)$ are some sort of count of some sort of particle, agent, or object across products. In general, one refers to such quantity as a ``mass'' occupying each product. For example, it could represent the number of people employed in the production of a product, and the elements of the stochastic matrix $\mtx{P}$ could represent probabilities of transitioning from a product to another product. In this equation, ``mass'' is conserved, so that $\sum_p n_p(t) = \sum_p n_p(t+1)$. This type of dynamics is called ``diffusion dynamics''.
\end{description}
Since $\mtx{C}$ is \emph{right}-stochastic (as opposed to $\mtx{P}$ which is left-stochastic), these two comments above apply for $\mtx{C}$ identically except swapping every ``left'' for every ``right'': multiplying on the right of $\mtx{C}$ represents a consensus dynamics, but multiplying on the left represents a diffusion process.\footnote{For a useful review of the main differences between diffusion and consensus dynamics, and how dynamical processes are constrained by the community structure on networks, see \citet{schaub2018structured}, and references therein.}


\subsection{Implications and insights}
\Cref{eq_relatedness} and \Cref{eq_imitation} represent a simple first approximation for the process of collective learning, supported by empirical observations. As we explained, the processes of relatedness and collective imitation are two types of processes that belong to the class of consensus dynamics \citep{degroot1974reaching}.\footnote{If $c$ producing $p$ is analogous to having a positive opinion (while not producing it is having a negative opinion), one can think of \Cref{eq_relatedness} as a process of consensus happening between products within places, and of \Cref{eq_imitation} as a process of consensus happening between countries for given products.} 

The paradigm of economic complexity tells us that the equation of collective learning across places and products is
\begin{align}
	\mtx{M}(t+1) &= \mtx{M}(t)\cdot \mtx{P} ~ + ~ \mtx{C}\cdot\mtx{M}(t).
\label{eq_collectivelearning}
\end{align}

These dynamics describe what drives diversification across most countries and most products, although not all of them. In particular, it describes the process of ``catching-up'' of places that are not fully diversified, but it does not explain the process of innovation that drives the production of completely new products by the most advanced economies. Hence, it is useful to remember that there may be an external driving force which we haven't talked about, which we will denote by $\mtx{U}(t)$. This force may as well be endogenous to the capabilities and products a places already has (diversification begets innovation). For now, we re-express our equation as:
\begin{align}
	\mtx{M}(t+1) &= \underbrace{\mtx{M}(t)\cdot \mtx{P}}_{\text{relatedness}} ~ + ~ \underbrace{\mtx{C}\cdot\mtx{M}(t)}_{\text{imitation}}  ~ + ~ \underbrace{\mtx{U}(t)}_{\text{innovation}}.
\end{align}

Equations similar to those of consensus dynamics are used in many ``recommender'' systems such as those at the base of platforms like Netflix and Amazon to suggest products to their customers. They are in effect predicting which items (i.e., ``products'') will be, most probably, watched/bought (i.e., ``produced'') by which users (i.e., ``countries''). This algorithmic approach is called ``collaborative filtering'' in the machine learning literature. It turns out we are doing the same in the framework of economic complexity.\footnote{The principles of relatedness and collective imitation can be combined, in principle, into a single term: $\mtx{M}(t+1) = \mtx{C}\cdot\mtx{M}(t)\cdot \mtx{P}$. This assumes some interaction between product relatedness and country similarity which we will not analyze here.}

Expressing the principle of relatedness and the principle of collective imitation in matrix form as in \Cref{eq_collectivelearning} reveals a few important mathematical properties that have economic and practical value. 

But before we comment on some of the implications, it must be said that these equations may not be the correct ultimate description of the process. In fact, many underlying micro-processes may give rise to the same macro-processes.\footnote{Think, as an analogy, of the Central Limit Theory in statistics: the mean of many random variables tends to be approximately normally distributed, regardless of the original statistical distribution of the random variables being averaged. In the same way, different mechanisms operating at the level of ideas, people and firms may give rise, effectively, to the same dynamical equations of ``collective consensus''.} These are questions that should be resolved by determining (i) \emph{what} is it that is flowing as places diversify (e.g., information vs. people vs. firms), (ii) what is the correct production function (i.e., in the last section we assumed a Leontief production function based on a multiplicity of capabilities, but this may be an extreme special case), and (iii) whether capabilities precede the output or the output precede the capabilities. At this point, however, we can analyze the consequences of describing the process of collective learning using general equations such as \Cref{eq_collectivelearning}.

As a first observation, simple matrix representations such as \Cref{eq_collectivelearning} suggest few ``low-dimensional'' quantities exist which can serve as summary statistics of the process. For us, then, it mean few quantities may exist that summarize the information of the whole process of collective learning. To understand why this is the case, we need to recall what ``eigenvalues'' and ``eigenvectors'' are. 

In general, any given matrix $\mtx{A}$ can be multiplied on the right by a column-vector $\vec{v}$, and that multiplication will result on another column-vector $\vec{w}=\mtx{A}\cdot\vec{v}$. However, there are some special vectors that when multiplied by the matrix $\mtx{A}$ just shrink or get expanded by a number, $\mtx{A}\cdot\vec{v} = a\vec{v}$, where $a$ is the shrinking/expanding factor. The matrix $\mtx{A}$ and all these multiplications have a physical meaning: $\mtx{v}$ is a point in space (a space which can have several dimensions), and the multiplication by $\mtx{A}$ moves the point somewhere else in the space. Hence $\mtx{A}$ describes in a sense the ``flow'' in a space because it determines how each point will move where. The eigenvectors point in directions where points just move to or away from the origin in a linear way.

For us, then, matrix $\mtx{P}$ determines how the vector of production of a specific country will ``move'' (it will diversify) in the ``Product Space'', while $\mtx{C}$ determines how a specific product will diffuse across countries, i.e., how the vector that tells us where the product is being produce will ``move'' in the ``Country Space''. The eigenvectors of these matrices will tell us the dominant axis or directions of movement.

Let us define the eigenvectors we need to describe these axis where countries and products are flowing into. 

Since $\mtx{M}(t)$ is multiplying $\mtx{P}$ on the left, we need $\mtx{P}$'s \emph{left}-eigenvectors:
\begin{align}
	\vec{\psi}_k^T \cdot\mtx{P} = \lambda_k \vec{\psi}_k^T,
\end{align}
where $\lambda_1 = 1 \geq \left|\lambda_k\right| \geq 0$ (this is a property of all stochastic matrices). Analogously for $\mtx{C}$, $\mtx{M}(t)$ is multiplying $\mtx{C}$ on the right, so we need $\mtx{C}$'s \emph{right}-eigenvectors:
\begin{align}
	\mtx{C}\cdot \vec{\varphi}_k = \gamma_k \vec{\varphi}_k,
\end{align}
where $\gamma_1 = 1 \geq \left|\gamma_k\right| \geq 0$.

To illustrate why eigenvalues are useful, let us first take a row vector of $\mtx{M}(t)$ representing the country $c$, and let's denote it by $\vec{m_c}(t)^T$. An element $p$ of this vector is just $M_{c,p}(t)$. Every vector can be represented as a linear combination of the eigenvectors of a matrix. Let us apply this decomposition, such that $\vec{m_c}(t)^T=\sum_k c_k(t) \vec{\psi}_k^T$, where $c(t)$ are just the linear coefficients multiplying the eigenvectors. Let us see what is going on when we multiply on the right by $\mtx{P}$:
\begin{align*}
	\vec{m_c}(t+1)^T &= \vec{m_c}(t)^T\cdot\mtx{P} \\
	&= \left(\sum_k c_k(t) \vec{\psi}_k^T \right)\cdot\mtx{P} \\
	&= \sum_k c_k(t) \vec{\psi}_k^T \cdot\mtx{P} \\
	&= \sum_k \lambda_k c_k(t) \vec{\psi}_k^T  \\
	&\approx \lambda_1 c_1(t) \vec{\psi}_1^T + \lambda_2 c_2(t) \vec{\psi}_2^T.
\end{align*}
The approximation comes from the (very!) important property that the eigenvalues $\lambda_k$, for $k\geq 2$, are (typically) in magnitude smaller than unity, so they increasingly shrink the eigenvector components of $\vec{m_c}(t)^T$ that are not aligned with the dominant (and subdominant) left-eigenvectors $\vec{\psi}_1^T$ and $\vec{\psi}_2^T$. Let us take a specific element $p$ of $\vec{m_c}(t+1)^T$ (i.e., a product):
\begin{align*}
	M_{c,p}(t+1) &\approx \lambda_1 a_1(t) \psi_{p,1} + \lambda_2 a_2(t) \psi_{p,2}.
\end{align*}

If we do the same but for a column vector $p$ of $\mtx{M}(t)$, $\vec{m_p}(t)$ that gets multiplied on the left by $\mtx{C}$, we arrive at:
\begin{align*}
	M_{c,p}(t+1) &\approx \gamma_1 p_1(t) \varphi_{c,1} + \gamma_2 p_2(t) \varphi_{c,2}.
\end{align*}

Put together, we have,
\begin{align}
	M_{c,p}(t+1) &\approx \lambda_1 c_1(t) \psi_{p,1} + \lambda_2 c_2(t) \psi_{p,2} + \gamma_1 p_1(t) \varphi_{c,1} + \gamma_2 p_2(t) \varphi_{c,2}.
\label{eq_finalapprox}
\end{align}

Some theorems and properties of stochastic matrices tell us that $\psi_{p,1}$ do not actually vary across products $p$, and thus $\psi_{p,1}\equiv 1, \forall p$.\footnote{Recall that $\mtx{P}$ is column-normalized. That is, $\vec{1}^T\cdot\mtx{P}=\vec{1}^T$. As can be seen, this equation, which expresses the fact that the values of each column of $\mtx{P}$ add-up to one, also expresses an eigenvalue-eigenvector relation. Specifically, since we know that the largest eigenvalue of a stochastic matrix is $\lambda_1=1$, the normalization equation also says that the left-eigenvector associated with the dominant eigenvalue $\lambda_1=1$ is the vector of ones. Hence, $\vec{\psi}_1^T = (1,1,\ldots,1)$.} Similarly, $\varphi_{c,1}\equiv 1$. Now, the coefficients of the eigenvector decompositions are really dot-products: $c_k(t)\equiv \vec{m_c}(t)^T\cdot\vec{\psi}_k$ are country $c$-specific variables, while $p_k(t)\equiv ^T\vec{\varphi}_k^T\cdot\vec{m_p}(t)$ are product $p$-specific variables. In particular, they tell us how aligned the rows/columns of $\mtx{M}(t)$ are with each of the corresponding eigenvectors. For example, $c_2(t) = \vec{m_c}(t)^T\cdot\vec{\psi}_2$ tells us how aligned is the country with the sub-dominant eigenvector of $\mtx{P}$, and $p_2(t) = \vec{\varphi}_2^T\cdot\vec{m_p}(t)$ how aligned is the presence of that product across countries with the sub-dominant eigenvector of $\mtx{C}$. Since we have assumed that $\mtx{M}(t)$ is a matrix of 0's and 1's, then ``aligned'' just means ``summed''. Thus, it can be seen that $c_1(t)$ is equal to the \emph{diversity} of the country, $d_c(t)$, while $p_1(t)$ is the \emph{ubiquity} of the product, $u_p(t)$. Moreover, $c_2(t)$ therefore is the sum of the elements $\psi_{p,2}$ of the products produced by country $c$, and $p_2(t)$ is the sum of the values of $\varphi_{c,2}$ of the countries where $p$ is produced.

\Cref{eq_collectivelearning} reduces to the following terms:
\begin{align*}
	M_{c,p}(t+1) &\approx \lambda_1 d_c(t) + \gamma_1 u_p(t) + \lambda_2 c_2(t) PCI_{p} + \gamma_2 p_2(t) ECI_{c},
\end{align*}
where we will refer $\psi_{p,2} = PCI_p$ as the ``Product Complexity Index'' value of product $p$, while $\varphi_{c,2} = ECI_p$ as the ``Economic Complexity Index'' value of country $c$. These indices reveal whether the production of a product $p$ by country $c$ is ``aligned'' according to what other countries are producing, and what other products the country is producing. But note how some products and some countries can induce the alignment more strongly on other countries, through the coefficients $c_2(t)$ and $p_2(t)$.\footnote{As we will see, depending on a specific definition of $\mtx{P}$ and $\mtx{C}$, we have that $ECI_c = \vec{m_c}(t)^T\cdot\vec{PCI}/d_c$, i.e., ECI of a country is the average PCI of the products it produces. This can simplify even further the equation into: $$M_{c,p}(t+1) = d_c(t) + u_p(t) + (\lambda_2 d_c(t)  + \gamma_2 u_p(t) )PCI_{p}ECI_{c}.$$} These coefficients are determined by the products and/or countries with high PCI and ECI, respectively. Can we expect these products and countries to be the most knowledge-intensive or knowledge-endowed? Under certain conditions, indeed we can, and this has been empirically demonstrated. The conditions under which this assumption holds is when the matrices of similarities between products and countries to have no clear clusters or communities, which is precisely the situation when the approximations we used only required considering the sub-dominant eigenvectors.

\citet{mealy2017new} recently examined in detail the interpretation of the original ECI and PCI (proposed in \citet{hausman2011atlas}) as indices to cluster countries and products in two groups. Among other things, they show that the reason ECI correlates with measures of economic performance is because it captures patterns of product specialization (some products induce economic more than others). In addition, \citet{mealy2017new} show mathematical connections to other dimensionality reduction techniques based on similarity matrices.

The general conclusion of looking at collective learning in low dimensions is that to a first approximation the appearance of a product $p$ in country $c$ in the next time-step, $M_{c,p}(t+1)$, is positively determined by the diversity of the country, the ubiquity of the product, the alignment of the production basket vector of $c$ with the vector of product complexities and the alignment of the presences of $p$ across countries with the vector of economic complexities.

\subsection{Does diffusion of inputs implies consensus of outputs?}
So far we have argued two main points: First, that economic development ought to be understood through the lens of a collective learning process because changes in the collective knowhow of a place have the largest effect on the probability of successful productive activities, as compared to improvements in technology or individual learning; Second, we have argued that the process of collective learning is the process of accumulation of capabilities, which manifests itself as a diversification process in the space of outputs. The important point about this is that the accumulation of capabilities occurs slowly and is enabled by the fact that products share capabilities and countries resemble each other. This whole process of collective learning can be described by a simple equation called the ``consensus dynamics'' equation. 

The first point was encoded in our model of economic complexity, in \Cref{sec_model}. A previous and simpler version of that model \citep{HidalgoHausmann2009,HausmannHidalgo2011} considered products and countries, and was stated as
\begin{align}
	\mtx{M}(t) = \mtx{{\mathcal C}}(t)\odot \mtx{{\mathcal P}}.
\label{eq_CcaPpa}
\end{align}
The matrix $\mtx{{\mathcal C}}(t)$ is a matrix of countries and the capabilities they are endowed with (which changes with time), while $\mtx{{\mathcal P}}$ is the matrix of products (as columns) and the capabilities required to produce them (which we assume are approximately constant). The operator $\odot$ is a ``Leontief operator'' such that a country produces a product only if the capabilities required by the product are a subset of the capabilities in the country.

The production process represented by \Cref{eq_CcaPpa} can be written using conventional matrix multiplication as follows:
$
	\mtx{M}(t) = \left\lfloor \mtx{{\mathcal C}}(t)\cdot (\mtx{{\mathcal P}}\cdot \mtx{{\mathcal A}}^{-1})\right\rfloor,
$
where $\left\lfloor ~x~\right\rfloor$ is the ``floor'' function that rounds down number $x$ to the largest integer less than $x$, and $\mtx{{\mathcal A}}$ is a diagonal matrix whose elements is the number of capabilities required per product (i.e., the capability-based complexity). The operation of rounding down is difficult to treat mathematically, but here it can in turn be re-expressed using some exponents. To see this, let us look at a specific element of this operation: $M_{c,p}(t) = \left\lfloor \sum_a {\mathcal C}_{c,a}{\mathcal P}_{a,p}/\sum_{a'} {\mathcal P}_{a',p}\right\rfloor$. Note how this sum is accumulating the fraction of capabilities $a$ required by $p$. It is only when the country has accumulated one hundred percent of the capabilities that the Leontief operator allows $c$ to produce $p$. One can assume a CES production function, and express this more generally as $\left(\sum_a {\mathcal C}_{c,a}^{\rho}{\mathcal P}_{a,p}/\sum_{a'} {\mathcal P}_{a',p}\right)^{1/\rho}$, where $-\infty < \rho < 1$, with $1/(1-\rho)$ being the elasticity of substitution between the factors of production ${\mathcal C}_{c,a}$. The CES formula, however, breaks down if we assume the production factors $C_{c,a}$ are binary. Hence, to model a range of production functions that go from arithmetic average to Leontief, it is more useful to write it as 
\begin{align}
	\mtx{M}(t) = \left( \mtx{{\mathcal C}}(t)\cdot (\mtx{{\mathcal P}}\cdot \mtx{{\mathcal A}}^{-1})\right)^\rho,
\label{eq_rhoexponentfunction}
\end{align}	
where $\rho$ now ranges from $1$ (arithmetic average which implies full substitutability between factors) to $\infty$ (Leontief function and zero substitutability).

If $\mtx{{\mathcal C}}(t)$ represents the counts of people employed across countries $c$'s and with capabilities $a$'s, one possible diffusion process could be written as
\begin{align}
	\mtx{{\mathcal C}}(t+1) = \mtx{F_{c'\leftarrow c}}\cdot \mtx{{\mathcal C}}(t) \cdot \mtx{F_{a\rightarrow a'}},
\label{eq_diffusionofcapabilities}
\end{align}
where we have assumed that the transition probability that takes people in the cell $(c,a)$ to the cell $(c', a')$ is separable $\Pr(c',a'|c,a)=\Pr(c'|c)\Pr(a'|a)$. In \Cref{eq_diffusionofcapabilities} we represent the probability $\Pr(c'|c)$ as the value $F_{c'\leftarrow c}$ and $\Pr(a'|a)$ as the value $F_{a\rightarrow a'}$. You can see that the matrix $\mtx{F_{c'\leftarrow c}}$ must be a left-stochastic matrix while $\mtx{F_{a\rightarrow a'}}$ a right-stochastic matrix.

A question to investigate is: what is the process that describes $\mtx{M}(t)$ and its changes in time if one assumes (i) a production function such as \Cref{eq_rhoexponentfunction}, and (ii) a diffusion process of the capabilities in the space of $\mtx{{\mathcal C}}(t)$ such as \Cref{eq_diffusionofcapabilities}? Do we retrieve the consensus dynamics that we proposed earlier?

The question may seem esoteric. But it goes to the heart of some of the assumptions of TEC: that individual learning is limited and that collective learning is mainly driven by accumulation of capabilities carried by individuals. In other words, the basic assumptions of TEC would suggest that since people are conserved, their capabilities diffuse. But at the level of collectives, what we observe is not diffusion but collective learning. Hence, we should be able to postulate a diffusion process that respects the conservation of ``mass'' (i.e., the number of brains in the system), \emph{and} a production function such that collective learning can be described by a process of related diversification and collective imitation. Currently, this is still an open question.


\section{Diffusion Maps}
In the previous sections, we used the fact that the eigenvalues of stochastic matrices are bound between $-1$ and $1$ to reach \Cref{eq_finalapprox} from \Cref{eq_collectivelearning}. Presently, we said that the process of collective learning could be described to a first order approximation by taking into account the effect only of the dominant and sub-dominant eigenvalues and their corresponding eigenvectors. The validity of such approximation, however, depends on the actual distribution of eigenvalues, and how fast they decay.

The approximations using the largest eigenvalues to represent economic diversification through few eigenvectors is a form of dimensionality reduction. This particular way of reducing dimensions is directly of the dynamics we have postulated about collective learning. These eigenspaces are called ``Diffusion Maps'' \citep{lafon_diffusion_2006,coifman_diffusion_2006}. Two major advantages emerge over traditional dimensionality reduction techniques (such as principal component analysis or classical multidimensional scaling): on the one hand, diffusion maps can account for nonlinear dependencies between observations, and the other, they preserve local structures. Since we have claimed that collective learning is a slow process and that, in addition, it is one that is structured by spaces of relatedness and similarity, we need a representation of the data that preserves the local structures, and therefore induces a local geometry. 

In what follows we will explore some of the implications from this approach (see also \citealp{mealy2017new} for the connection between diffusion maps and other dimensionality reduction techniques in the context of TEC).

\subsection{Consensus dynamics and clustering}
The distribution of eigenvalues, the so called ``spectrum'' of a matrix, is determined by the number of communities in the matrix of similarities. If the nodes of the network are organized in well-defined $K$ clusters, then there are $K-1$ relatively large, nontrivial, eigenvalues, in addition to the dominant eigenvalue with value equal to 1. Thus, a heuristic that can be used to infer the number of communities in a network is to count the number of eigenvalues before we observe a large ``gap'' between pairs of consecutive eigenvalues. In practice, one can take the values larger than 0.1.\footnote{In physics, one compares the distribution of observed eigenvalues with those eigenvalues of a random matrix. For a few special cases of random matrices there are closed analytical formulas. This branch of physics is called Random Matrix Theory.} Second, the eigenvectors (both left- and right-eigenvectors) associated with those $K-1$ nontrivial eigenvalues reveal the structure of the clusters. Hence, if the clusters are well-defined, even carrying out a simple $K$-means clustering on the matrix $\mtx{\Phi}_{C\times K-1}$ where the columns are the $K-1$ right-eigenvectors of $\mtx{C}$  would identify the $K$ clusters (or, if one is interested in the products, one takes $\mtx{\Psi}_{P\times K-1}$ where the columns are the $K-1$ left-eigenvectors of $\mtx{P}$).\footnote{The Economic Complexity Index (ECI) is just one of these dimensions.} Let us see mathematically how this works. 

Recall that the principle of collective imitation posited that there exists a similarity matrix $\chi(c, c')$ that tells us the \emph{direct} pairwise influence that country $c'$ has on country $c$. That is, if country $c'$ produces one unit of product $p$, then this will have a direct influence on $c$ such that it will add $\chi(c, c')/\sum_{c''}\chi(c, c'')$ units to its production of $p$. Since $c'$ may influence another country $c''$, and $c''$ in turn may influence $c$, then the net influence of $c'$ on $c$ can be higher than its direct influence. We want to represent the process of collective imitation in low dimensions, and this requires us to define a metric that takes into account the full connectivity of the points defined by $\chi(c, c')$.

The framework of diffusion maps requires two assumptions: (i) symmetry $\chi(c, c') = \chi(c', c)$ and (ii) pointwise positivity $\chi(c, c')\geq 0$ for all $c$ and $c'$. Given the matrix of similarities $\chi(\cdot, \cdot)$, the idea is to state that two points (say, countries) $c$ and $c'$ should be considered to be ``close'' not only if $\chi(c, c')$ is high but, more generally, if they are connected by many short paths in the network defined by $\chi(\cdot, \cdot)$. 

Let $d_c = \sum_{c'}\chi(c, c')$, and let $\pi_c = d_c/\sum_{c'}d_{c'}$. From the symmetry property we get that $\pi_c C(c, c')=\pi_{c'} C(c', c)$. Therefore, the symmetry of $\chi(c, c')$ is important because it implies that $C(c, c')=\chi(c, c')/\sum_{c''}\chi(c, c'')$ is a (right) stochastic matrix which defines a \emph{reversible} Markov chain, which in turn implies nice properties about its eigenvalue/eigenvector decomposition, $$\mtx{C}\cdot \vec{\varphi}_k = \gamma_k\vec{\varphi}_k$$ and $$\vec{\xi}_k^T\cdot\mtx{C} = \gamma_k\vec{\xi}_k^T,$$ where $1=\gamma_1>|\gamma_2|\geq |\gamma_3| \geq \ldots \geq 0$. We normalize the eigenvectors such that $\sum_c \xi_{c,k}^2/\xi_{c,1} = 1$ and $\sum_c \varphi_{c,k}^2\xi_{c,1} = 1$. Eigenvectors are related according to $$\varphi_{c,l}=\frac{\xi_{c,l}}{\xi_{c,1}},\quad\text{for all $c$}.$$
Thus, we have that left and right eigenvectors are orthonormal $\vec{\xi}_k^T\cdot\vec{\varphi}_l = \delta_{k,l}$.

As has been explained previously, multiplying on the left of a right-stochastic matrix propagates probabilities. For us, this is simply a mathematical abstraction about a random diffusion process defined by $\mtx{C}$. According to such a process, the probability of a random walker of being across different nodes (i.e., countries $c$) at a time $t$ given an initial probability $\vec{\pi}(t=1)^T=\vec{\pi}_1^T$ is $\vec{\pi}(t+1)^T = \vec{\pi}_1^T\cdot \mtx{C}^t$. We have that the long-term stationary distribution is directly given by the vector $\lim_{t\rightarrow\infty}\vec{\pi}(t)^T =\vec{\xi}_1^T=\vec{\pi}^T$ (i.e., the left-eigenvector associated with the eigenvalue $\gamma_1 = 1$ is the stationary distribution and is, at the same time, proportional to the diversity of countries). 

Let us define the ``diffusion distance'' \citep{coifman2005geometric} as follows:
\begin{align}
	D_t^2(c,c') &= \left\|C(c,\cdot) - C(c',\cdot)\right\|^2_{1/{\pi}} \nonumber\\
	&= \sum_{c''}\frac{\left(C_t(c,c'') - C_t(c',c'')\right)^2}{\pi},
\label{eq_defdiffusiondistance}
\end{align}
where $C_t(c,c')$ is an element of the matrix $\mtx{C}^t$. As can be seen, this type of distance adds up all the contributions from several paths relating $c$ and $c'$ and, as a consequence, is robust to noise in the specific measurements of $\chi(\cdot,\cdot)$.

The spectral decomposition of $C_t(c,c')$ is
\begin{align}
	C_t(c,c') = \sum_k \gamma_k^t \varphi_{c,k} \xi_{c',k}.
\end{align}
Replacing this spectral representation in \Cref{eq_defdiffusiondistance} of the diffusion distance yields
\begin{align}
	D_t^2(c,c') &= \sum_{k\geq 2} \gamma_k^{2t} \left(\varphi_{c,k} - \varphi_{c',k}\right)^2.
\label{eq_diffdisteig}
\end{align}
Note that the term for $k=1$ disappears from the sum because $\vec{\varphi}_1 = \vec{1}$. 

If there are $K$ communities or clusters of countries, there will be $K$ significantly large eigenvalues. If that is the case, \Cref{eq_diffdisteig} can be approximated using only $K-1$ terms:
\begin{align}
	D_t^2(c,c') &\approx \sum_{k = 2}^{K} \gamma_k^{2t} \left(\varphi_{c,k} - \varphi_{c',k}\right)^2 \nonumber\\
	&=\left\| \vec{\Phi}_t(c) - \vec{\Phi}_t(c) \right\|^2,
\label{eq_diffdisteigapprox}
\end{align}
where $\vec{\Phi}_t(c)$ is a the vector of coordinates of a point $c$ located in a euclidean space of $K-1$ dimensions, where a given coordinate $k$ is $\Phi_{t,k}(c) = \gamma_k^{t}\varphi_{c,k}$.

We proposed that collective learning can be described by consensus dynamics, and we use stochastic matrices to express the mathematical equations of these dynamics. In this framework, right-eigenvectors and left-eigenvectors contain information about the community structure of these networks of similarities. However, the framework of diffusion maps reveals that right-eigenvectors (in the case of $\mtx{C}$, and left-eigenvector for $\mtx{P}$) are better at capturing the information about communities. Hence, countries in the right-eigenspace will flock together, and their movement in this euclidean space will align more strongly with that of other countries in their community \citep{coifman_diffusion_2014}. (We will show later that, in contrast, the left-eigenvectors are better for measuring capabilities such that it is the distance to the origin that will be related to the underlying number of capabilities). These two spaces (the left and right eigenspaces) are related since the extent to which a country is embedded in a community is itself a measure of the number of capabilities it has.\footnote{A way to see this is by recalling Anna Karenina's Principle: Richly diversified countries tend to be all alike, while poorly diversified countries are poorly diversified in their own way. Hence, we should expect a ``club of rich countries'' to emerge.}

To get a grasp for how this framework is applied, we create three $\mtx{M}$ matrices. The first, we defined some set of countries by the products they produce. In this way, we create regions of countries that specialize in certain products. Thus, we created the matrix by putting $M_{c,p}=1$ with a probability of 0.6 if $c$ and $p$ belong to the same community, and with probability 0.1 if not. The second way is following the model of \citet{HausmannHidalgo2011}, using $\mtx{{\mathcal C}}$ and $\mtx{{\mathcal P}}$ to determine $\mtx{M}$. These underlying matrices are also binary matrices, which can be thought of as the matrix of countries and the capabilities they have on the one hand, and the matrix of products and the capabilities they require to be produced on the other, such that $\mtx{M}=\mtx{{\mathcal C}} \odot \mtx{{\mathcal P}}$. Finally, the third way to construct $\mtx{M}$ is from real data. We choose the year 2015, 224 countries and 773 products (SITC4 codes).

\begin{figure}[h]
\begin{center}
\includegraphics[width=\textwidth]{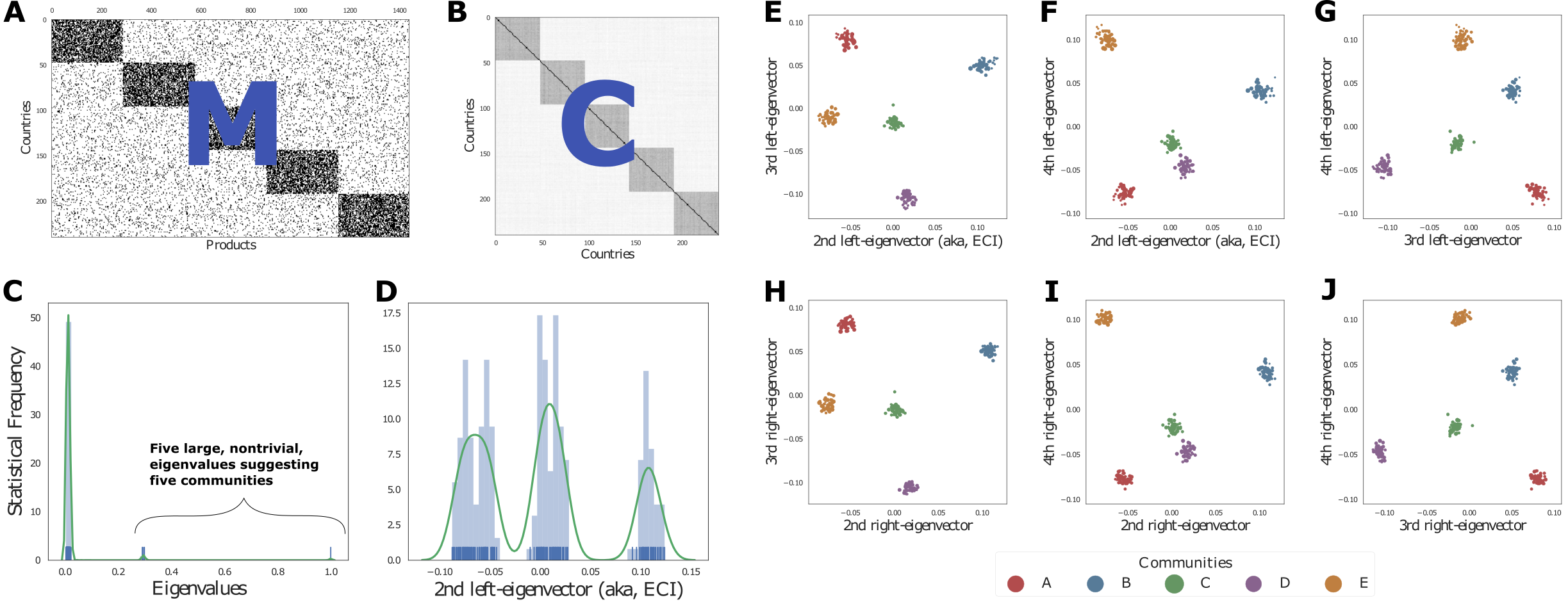}
\caption{{Example of a matrix connecting countries with products with a uniform
probability. The within-community probability was set at 0.6 and the
between-community probability at 0.1.
{\label{340547}}%
}}
\end{center}
\end{figure}

\begin{figure}[h]
\begin{center}
\includegraphics[width=\textwidth]{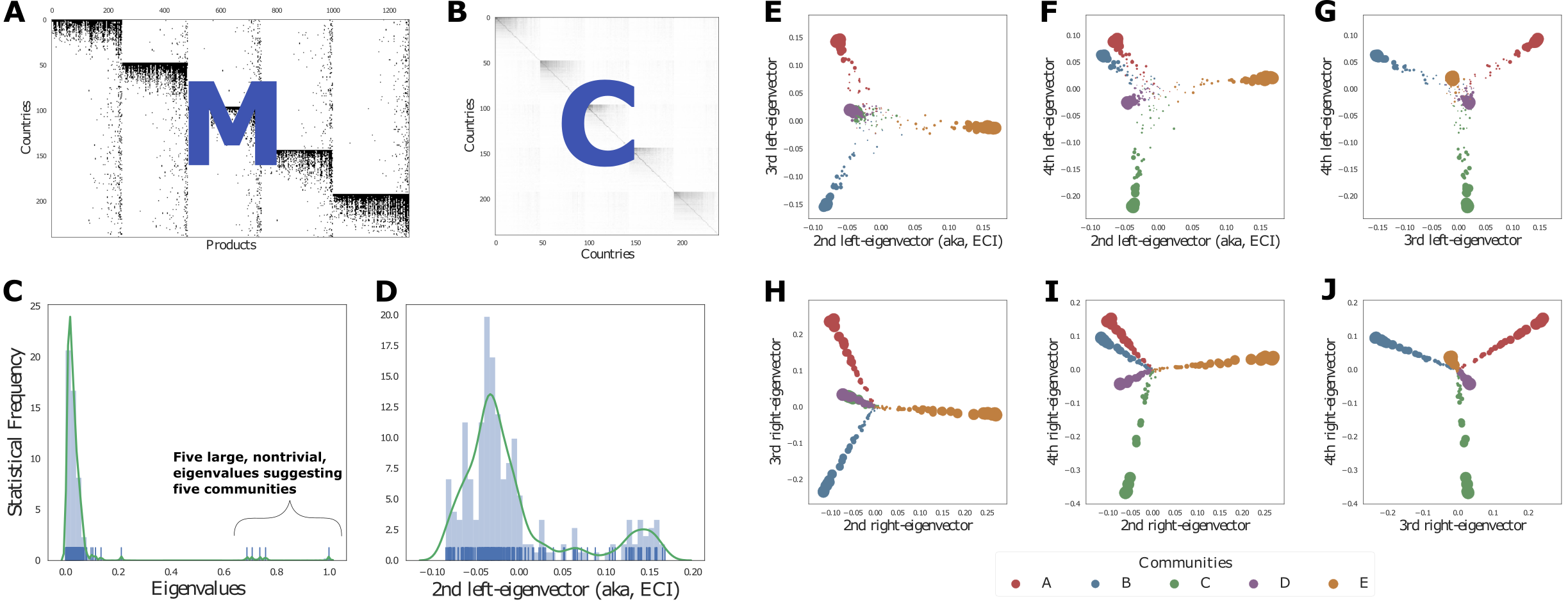}
\caption{{Example of a matrix connecting countries with products as it results
from the interaction between the matrix of countries and capabilities,
with the matrix of products and the capabilities required. Within each
community we model a nested pattern in which some countries have many
capabilities and others only a few. We also include the possibility in
which some products can be produced countries regardless of the
community to which they belong.~
{\label{369902}}%
}}
\end{center}
\end{figure}

\begin{figure}[h]
\begin{center}
\includegraphics[width=\textwidth]{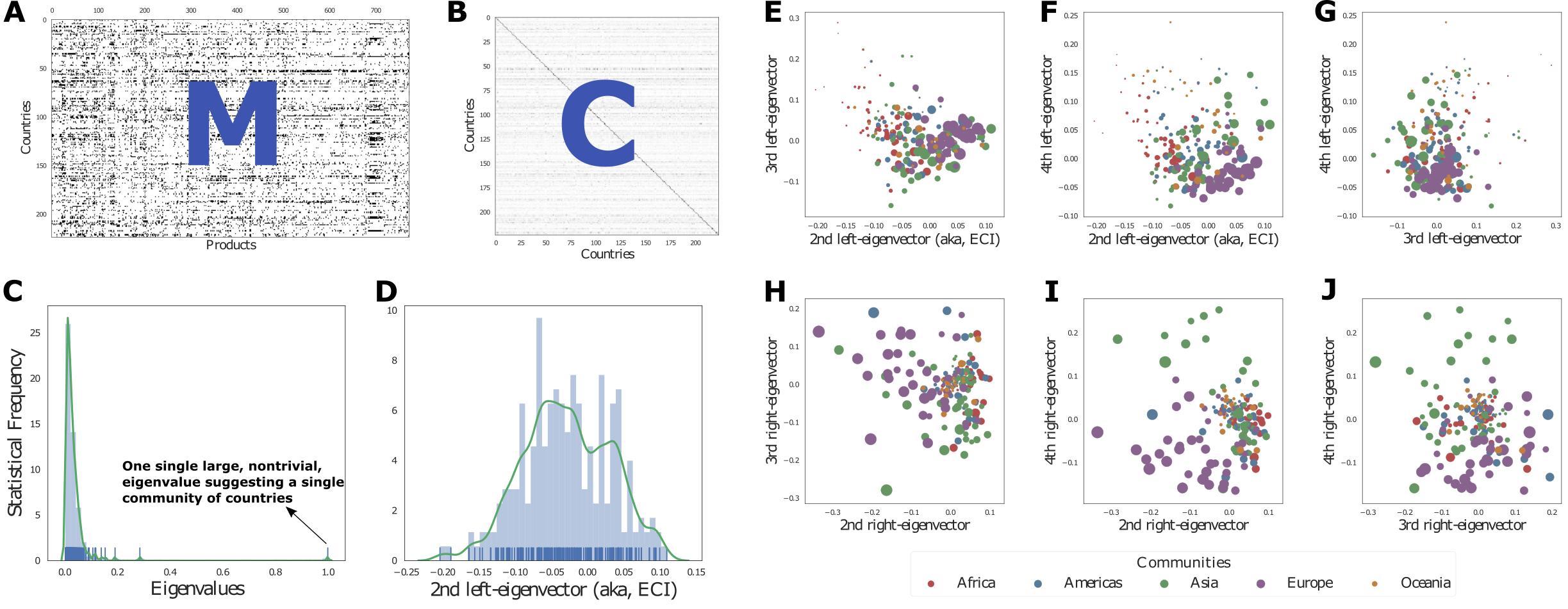}
\caption{{The same exercise as in Figures 1 and 2, but with real data from 2015
using only exporters and their products (see
Fig.~{\ref{660559}} for exporters vs.
importer-products). As can be seen, communities are less clear in this
representation.
{\label{583738}}%
}}
\end{center}
\end{figure}

Figure~{\ref{340547}} shows the results from the matrix
filled uniformly, with five communities.
Figure~{\ref{369902}} ~shows the results from the
matrix created based on an underlying structure of capabilities, with
also five communities. Figure~{\ref{583738}} show the
results from real data.

\Cref{583738} does not reveal any special structure. In order to see more structure, let us disentangle technology from geography. The goal here is to represent the process of collective imitation, but taking into account \emph{geographical communities}. In other words, we want to model the fact that countries in a same geographical region tend to export to the same importers. We accomplish this by adding the dimension of the importers to the matrix $\mtx{M}$.\footnote{Michele Coscia was who originally saw this, and made the connection between ECI and clustering by considering the geographical dimension of importers.} Hence, we will work below with the information $M_{c,i,p}$, where $c$ is the index of the exporter country, $i$ the importer, and $p$ the product traded. We will ``flatten'' (or ``widen'') this three-dimensional matrix such that we make it a two-dimensional matrix with a very wide set of columns $M_{c,(i-p)}$, where rows are the exporters $c$, but each column is a importer-product combination.

The following figure (Fig. {\ref{660559}}) shows this matrix. As can be observed, it 
also displays a triangular pattern. We construct its corresponding $\mtx{C}$ matrix, and compute the eigenvalues to
have an indication of how many clusters there are. Finally, we plot the
3-dimensional left-eigenspace (top-row of figures in the right 6-panel
plot), and the 3-dimensional right-eigenspace. 
\begin{figure*}[h]
	\begin{center}
		\includegraphics[width=0.99\textwidth]{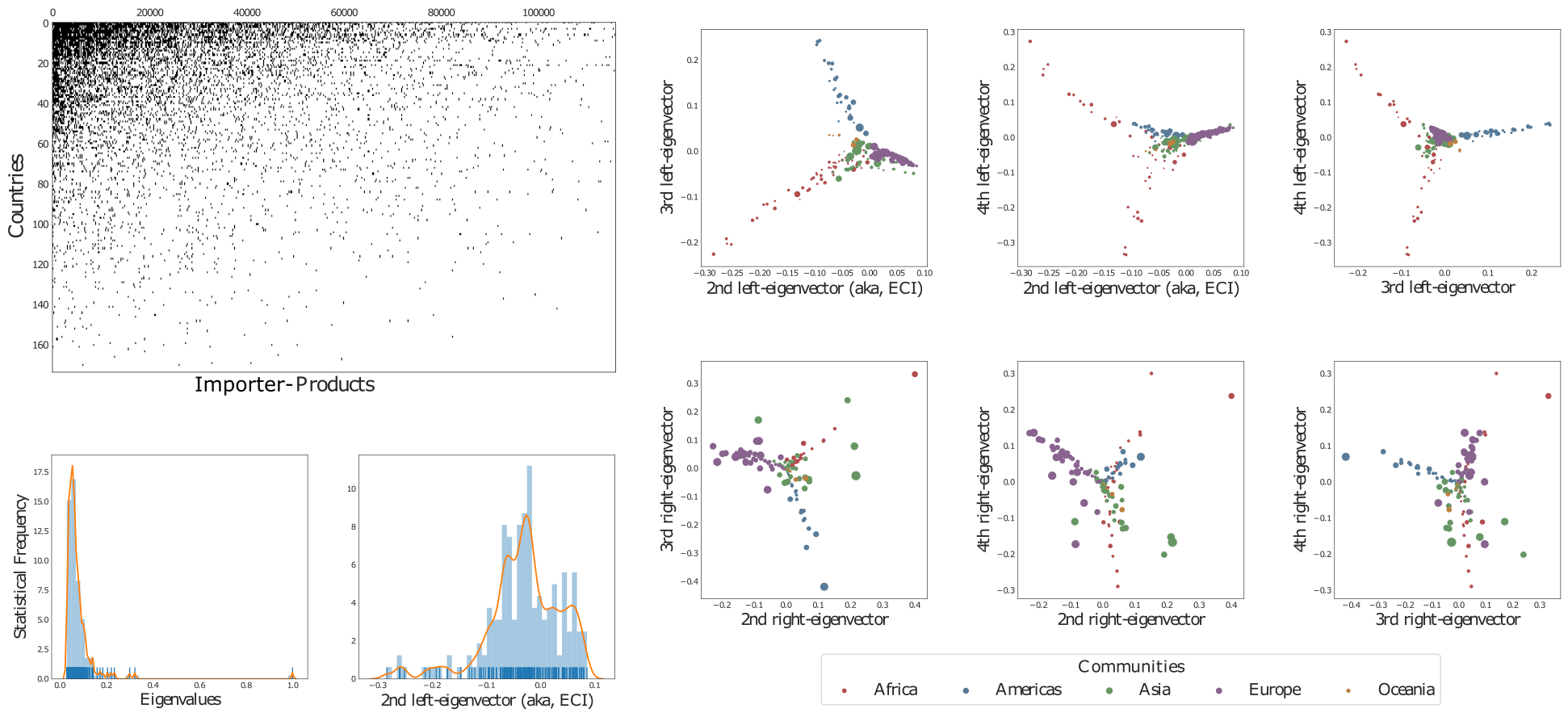}
	\caption{{Real data from the matrix of exporters (rows) vs importer-products
	(columns). The density of eigenvalues provides a sense of the number of
	communities of exporters by counting the number of large eigenvalues.
	The density of the distribution of values of ECI is a further indication
	of the number of communities. However, the communities become clear on
	the left-eigenspace (each dot is an exporter), which is shown on the
	three row-panels. We color the five continents of the exporters,
	supporting the idea that exporters belong to the same geographical
	communities. The bottom row of 3 panels shows the right-eigenspace,
	which we hypothesize provides a measure of the underlying capabilities
	of exporters.
	{\label{660559}}%
	}}
	\end{center}
\end{figure*}

The meaning of these clusters of countries, in the context of consensus dynamics, is the following: diversification will occur first within the communities, and then globally. Once a community has reached consensus, the countries within the community will ``flock'' together in the same way.


\newpage
\section{Methodological notes}


\subsection{Some comments about processing real data}
Studying the process of collective learning at the level of countries has some benefits in contrast to studying it at the level of cities. In particular, by studying trade between countries, one can look at the capacity of different societies to produce products and compete at a worldwide scale. Different countries rely on different institutions, have different laws, and political systems. But since the World is big, competition is fierce and the number of potential buyers is so large, the ability to produce something competitively is an accurate indication of the number of capabilities that a country has. 

To identify capabilities in a place, and study how capabilities flow across countries, or how countries acquire the ability to diversify, it would be useful to know what a place produces consistently, significantly and systematically. In brief, we should identify if a place $c$ is producing a product $p$ ``competitively''. How? 

The fundamental idea is to compare the \emph{observed} production against the \emph{expected}:
\begin{align}
	\text{Measure of competitiveness}&=\frac{\text{Observed production}}{\text{Expected production}}.
\end{align}
Mathematically, we write this as
\begin{align}
	R_{c,p} = \frac{X_{c,p}}{\E{X_{c,p}}},
\label{eq_observedexpected}
\end{align}
where $\E{\cdot}$ is the expectation operator. Every expectation assumes, explicitly or implicitly, a model of the world. To answer how competitive a place is, we need to define a simple model of the world, a ``null model''. 

A specific null model is to assume that countries should be expected to produce a product in the same proportion to its total output, as the share of that product of total worldwide production: $$\E{X_{c,p}} = X_c \left(\frac{X_p}{X_{\text{total}}}\right).$$

This defines what is known as the measure of Revealed Comparative Advantage (or Location Quotient in the context of urban economics and regional science). If $X_{c,p}$ represents the export value of country $c$ in product $p$:
\begin{align}
	R_{c,p}\equiv \frac{X_{c,p}/\sum_p X_{c,p}}{\sum_c X_{c,p}/\sum_{c,p} X_{c,p}}.
\label{eq_rcadef}
\end{align}
This can be computed using matrix operations as $\mtx{R} = X_T (\mtx{x_c}^{-1}\cdot \mtx{X} \cdot \mtx{x_p}^{-1})$, where $X_T$ is the total sum of values of matrix $\mtx{X}$, $\mtx{x_c}$ is a square diagonal matrix whose elements are the total exports by country, and $\mtx{x_p}$ is a square diagonal matrix whose elements are the total exports by product.

The values of $X_{c,p}$ are typically heavy-tailed distributed. That is to say, their magnitude can range across several orders of magnitude. Recall, in particular, that our null model predicts lognormally distributed measures of output (see discussion in \Cref{sec_multadd}). The divisions and multiplications in the particular definition of \Cref{eq_rcadef} compound this variability, and $R_{c,p}$ have values that are sometimes even more skewed and extreme. However, one is typically more interested in characterizing the competitiveness of places with regard to their production processes using quantities that behave more ``mildly''. So how does one ``tame'' the values of matrices such as $R_{c,p}$?

One way is to simplify matters and create $M_{c,p} = 1$ if $R_{c,p}>1$ and 0 otherwise. This binarization has several motivations, one of which is reducing the noise. The threshold $R_{c,p}=1$ is natural, since it separates the ``more-than-expected'' from the ``less-than-expected'' values. However, this operation is also hiding potentially important information contained in the specific variations below or above that threshold, not to mention the fact that some values may be larger than the threshold but may not actually be statistically significant.

A natural transformation is thus to take logarithms. The problem is that many values of $X_{c,p}$, and consequently of $R_{c,p}$, are zero. Directly applying logarithms is thus not appropriate. Adding a 1 before taking logarithms, although often done, $\mathrm{log1p}(R_{c,p})\equiv \log(R_{c,p}+1)$, is also not appropriate because adding 1 artificially creates a characteristic scale in a variable that, given its broad statistical distribution, is in fact better described by a ``scale-free'' distribution.\footnote{What is really happening is that $R_{c,p}$ has an intrinsic characteristic scale that is approximately equal to 1, but its variation is multiplicative. Adding a 1 completely changes the distribution, which is why $\log(1+R_{c,p})$ is \emph{not} a normally distributed random variable.} One proposal is thus to implement the following piecewise function:
\begin{align}
	\widetilde{R}_{c,p}=
	\begin{cases}
		0&,\text{ if $R_{c,p}=0$,} \\
		1+\left(\frac{r_0-1}{\log(r_0)}\right)\log(R_{c,p})&,\text{ otherwise,}
	\end{cases}
\end{align}
where $r_0\equiv \min_{c,p}\left(R_{c,p}|R_{c,p}>0\right)$. This transformation is useful because it maps 
\begin{eqnarray*}
	R_{c,p}=r_0&\longleftrightarrow&\widetilde{R}_{c,p}=r_0, \\
	R_{c,p}=0&\longleftrightarrow&\widetilde{R}_{c,p}=0, \\
	R_{c,p}=1&\longleftrightarrow&\widetilde{R}_{c,p}=1. 
\end{eqnarray*}
Thus, it respects the usual bounds of $R_{c,p}$ but is distributed approximately normal.

There is a more agnostic and general way of computing measures of ``competitiveness''. It is simply the observation that \Cref{eq_observedexpected} is attempting to estimate a residual. Thus, one can simply create a regression model (linear or non-linear, depending on the needs), and retrieve the residuals of such regression. The natural threshold to separate the ``more-than-expected'' from the ``less-than-expected'' is 0. For example, one can generalize \Cref{eq_rcadef} as
\begin{align}
	R_{c,p} = y_{c,p} - \widehat{y_{c,p}}, 
\end{align}
where $y_{c,p}=\log(X_{c,p}+1)$ and $\widehat{y_{c,p}}$ is the OLS estimate that minimizes the squared error according to a model $$y_{c,p} = \beta_0 + \beta_1\log(\sum_p X_{c,p}) + \beta_2\log(\sum_c X_{c,p}) + \varepsilon_{c,p}.$$ Note that we \emph{can} add a 1 to $X_{c,p}$ because the characteristic scales of $X_{c,p}$ are so much larger than 1. Residuals from these type of regressions, $R_{c,p}\equiv \widehat{\varepsilon_{c,p}}$, can be generalized even further if one wants to complicate the model and add variables to control for population size, geography, or presence of natural resources (to name a few possibilities).


\subsection{Some comments about Product/Country Spaces of Relatedness/Similarity}
Recall that the principles of relatedness and the collective imitation assume there are some matrices of similarities. Where do we get those similarity matrices, or how do we construct them?

Here are some possibilities and notes about ways to compute the relatedness of products, which we encode in a similarity matrix relating every pair $p$ and $p'$. The same concepts could apply to the matrix of country-country similarities.
\begin{itemize}
	\item Hidalgo et al. (2007):
	\begin{align}
		\phi(p, p') = \min\{ \Pr(p|p'), \Pr(p'|p) \}
	\end{align}
	where $\Pr(p|p')$ can be set to mean the fraction of countries for which $M_{c,p}=1$ \emph{given} they also have $M_{c,p'}=1$.
	
	\item More generally, one could simply attempt a variety of similarity measures between columns (see, for a review of measures, \citealp{cha2007comprehensive}). For example, correlate the columns of the matrix $\widetilde{R_{c,p}}$:
	\begin{align}
		\phi(p, p') = \mathrm{cor}( \vec{\widetilde{R}}(p), \vec{\widetilde{R}}(p') ).
	\end{align}
	
	\item In general, one is typically trying to infer whether there is \emph{any} statistical dependency between the production of products. A correlation only measures \emph{linear} dependency. Hence, the question is whether $\Pr(p,p')$ is just the product of the marginal probabilities $\Pr(p)\Pr(p')$ (which would imply independence). If $\Pr(p,p')$ is larger, then the products are positively dependent, if it is less, the products are negatively dependent. This suggests a measure of dependence called a ``pointwise mutual information'' (one measure of particular interest would be \citealp{reshef2011detecting}):
	\begin{align}
		\phi(p, p') = \log_2\left(\frac{\Pr(p,p')}{\Pr(p)\Pr(p')}\right).
	\end{align}
	
\end{itemize}

We can take other more ``formulaic'' approaches. Thus, the original product space can be ``un-packed'' mathematically as:
\begin{align}
	\Pr(p|p') &= \frac{\Pr(p,p')}{\Pr(p')} \nonumber \\
	&= \frac{N(p,p')/N_c}{N(p')/N_c} \nonumber \\
	&= \frac{\sum_c M_{c,p}M_{c,p'}}{\sum_c M_{c,p'}}  \nonumber \\
	&= \frac{\sum_c M_{c,p}M_{c,p'}}{u(p')}. 
\end{align}
Symmetric matrix:
\begin{align*}
	\mtx{\Phi} &= \min\left\{\mtx{U^{-1}}\cdot\mtx{M}^T\cdot\mtx{M}~,~~~~ \mtx{M}^T\cdot\mtx{M}\cdot\mtx{U^{-1}}\right\}.
\end{align*}
Other options can be seen as generalizations of these combinations of matrices\footnote{Alje van Dam and Koen Frenken have work-in-progress linking and generalizing measures of similarity}:
\begin{itemize}
	\item $\mtx{\Phi} = \mtx{M}^T\mtx{M}/N_c$ (symmetric [joint freq.])
	\item $\mtx{\Phi} = \mtx{M}^T\cdot\mtx{D^{-1}}\cdot\mtx{M}$ (symmetric [``diversity normalized'' joint freq.])
	\item $\mtx{\Phi} = \mtx{M}^T\cdot\mtx{D^{-2}}\cdot\mtx{M}$ (symmetric [``averaged probabilities''])
	\item $\mtx{\Phi} = \mtx{U^{-1}}\cdot\mtx{M}^T\cdot\mtx{M}\cdot\mtx{U^{-1}}$ (symmetric [mutual information])
\end{itemize}

\subsection{Conventional calculation of ECI}
\label{sec:ECIdefinition}
The actual calculation of the ECI uses the similarity $\mtx{M}\cdot\mtx{U^{-1}}\cdot\mtx{M}^T$, where $\mtx{U}$ is matrix whose diagonal values are the ubiquities of products. Let us see why we use this specification. 

Let  matrix $\mtx{M}$ have size $C\times P$. This is a matrix that has been discretized so that $M_{c,p}$ is 1 if the product $p$ is exported in country $c$, and 0 otherwise. From this matrix, one creates two stochastic matrices. First, the right-stochastic (i.e., row-stochastic or row-normalized) transition matrix of ``countries to products'', $$\mtx{R}=\mtx{D}^{-1}\cdot\mtx{M},$$ and second, the left-stochastic (i.e., column-stochastic or column-normalized) transition matrix of ``products to countries'', $$\mtx{L}=\mtx{M}\cdot\mtx{U}^{-1},$$ where $\mtx{D}$ is the matrix whose diagonal elements are the diversities of countries, $\vec{d}=\mtx{M}\cdot\vec{1}$ and similarly for $\mtx{U}$, whose diagonal is $\vec{u}=\mtx{M}^T\cdot\vec{1}$, the vector that contains the number of countries from which the product is exported (i.e., its \emph{ubiquity}). We use the notation $\diag{\vec{x}}$ to mean the matrix whose diagonal is the vector $\vec{x}$ and the other values are zero, and $\vec{1}$ to denote a vector of 1's.

Two comments:
\begin{itemize}
	\item The matrix $\mtx{R}$ takes averages of when multiplied by a vector on the \emph{right}. Consider $\vec{y}$ a vector in which each element is a property of each product. 
	Then $\mtx{R}\cdot \vec{y}$ is that average value of the property per country.
	\item The matrix $\mtx{L}$ takes averages of when multiplied by a vector on the \emph{left}. This time, consider another general vector $\vec{x}^T$ in which each element is a property of each country. Then $\vec{x}^T\cdot\mtx{L}$ is that average value of the property per product.
\end{itemize}
Recall that we are using ``stochastic'' matrices, not because we are modeling a stochastic process, but because of the principles of relatedness and collective imitation, which are based on averaging.

Let us construct the left-stochastic transition probability matrix of ``countries to countries'', $$\mtx{C}=\mtx{R}\cdot\mtx{L}^T=\mtx{D}^{-1}\cdot\mtx{M}\cdot\mtx{U}^{-1}\cdot\mtx{M}^T.$$ As can be seen, this can be written as $\mtx{C} = \mtx{D}^{-1}\mtx{S}$, where $\mtx{S}$ has the elements of country-country similarity $\chi(c,c')$. Thus, this is one possible version of the matrix implied in \Cref{eq_imitation}.

For mathematical convenience, we will assume that the stochastic matrix $\mtx{C}$ is irreducible and aperiodic.\footnote{When $\mtx{C}$ is constructed using real data, it is irreducible since all countries produce at least one product that some other country also produces, and it is aperiodic since, by construction, it has self-loops.} 

Now, let $\vec{l_i}^T$ and $\vec{r_i}$ be the $i$th left-eigenvector and right-eigenvector, respectively, so that the eigenvalues are ordered in decreasing value, $1=\gamma_1\geq \gamma_2\geq\cdots\geq \gamma_C$. The list of ECIs for countries is defined as the right sub-dominant eigenvector, $\vec{ECI}\equiv \vec{r_2}$:
\begin{align}
	\mtx{C}\cdot\vec{ECI}=\lambda_2 \vec{ECI}.
\label{eqn:drag}
\end{align}
Remember, we take the right-eigenvector because the matrix $\mtx{C}$ is multiplied on the right by $\mtx{M}$, which represents the phenomenon in which the production of products across countries changes according to the principle of collective imitation.

It is easy to prove that the vector $\vec{d}$ of the diversity of countries is orthogonal to the vector of ECIs, $\vec{ECI}$, once you realize that $\vec{d}$ is actually the dominant left-eigenvector (sometimes referred to as the ``perron'' eigenvector, or just simply, the stationary distribution of the discrete markov chain defined by $\mtx{C}$). Thus, multiplying $\vec{d}$ on the left of $\mtx{C}$, and expanding $\mtx{C}$ into its components,
\begin{align}
    \vec{d}^T\cdot\mtx{C} &= \vec{d}^T\cdot(\mtx{R}\cdot\mtx{L}^T), \nonumber \\
    &= \vec{d}^T\cdot(\mtx{D}^{-1}\cdot\mtx{M}\cdot\mtx{U}^{-1}\cdot\mtx{M}^T), \nonumber \\
    &= \vec{d}^T\cdot(\diag{1/\vec{d}}\cdot \mtx{M})\cdot(\diag{1/\vec{u}}\cdot\mtx{M}^T), \nonumber \\
    &= (\vec{1}^T\cdot\mtx{M})\cdot(\diag{1/\vec{u}}\cdot\mtx{M}^T), \nonumber \\
    &= \vec{u}^T\cdot(\diag{1/\vec{u}}\cdot\mtx{M}^T), \nonumber \\
    &= \vec{1}^T\cdot\mtx{M}^T, \nonumber \\
		&= (\mtx{M}\cdot\vec{1})^T, \nonumber\\
    &= \vec{d}^T.
\end{align}
Thus, $\vec{d}^T$ is a left-eigenvector of $\mtx{C}$ associated with the eigenvalue $\gamma_1=1$, which (from the Perron-Frobenius theorem) one concludes that $\vec{d}^T$ is the \emph{dominant} left-eigenvector. This means, given classical results from discrete markov chains, that the stationary distribution of the stochastic process defined by $\mtx{C}$ is ${\boldsymbol \pi} = \vec{d}/\sum_c d_c$. Therefore, since left-eigenvectors are orthogonal to right-eigenvectors, $\vec{l_i}^T \cdot \vec{r_j}\propto\delta_{i,j}$, we conclude that $$\vec{d}^T\cdot\vec{ECI}=0,$$ which is a result that had been noted before already by \citet{kemp2014interpretation}.

All these same results apply to the product space matrix, $\mtx{P} = \mtx{R}^T\cdot \mtx{L}$, except all ``left'''s are swapped with ``right'''s. Namely, the sub-dominant left-eigenvector is the list of product complexity indices, PCIs, and the dominant right-eigenvector is proportional to the list of ubiquities.

In the literature it is sometimes said that the economic complexity index can be axiomatically defined by postulating that products have a complexity, that the complexity of countries is the average complexity of the products it exports, and that the complexity of the products as the average complexity of the countries where it is exported. It is claimed that this \emph{uniquely} defines these two vectors. In other words, the claim is that $\vec{x}^T\cdot\mtx{L}=\vec{y}^T$ and $\mtx{R}\cdot\vec{y}=\vec{x}$ uniquely define the vectors $\vec{x}^T$ and $\vec{y}$, and that these correspond to the economic complexity of countries and products, respectively. This is not true, however. Any pair of the right/left-eigenvectors of the matrices $\mtx{C}$ and $\mtx{P}$ have this precise property, and thus this does not uniquely define what ``complexity'' means.\footnote{There an additional complication arising from the fact that $\mtx{M}$ is not square. In general, $C$ is less than $P$, and this implies that $P$ can have, at most, $C$ linearly independent columns, which in turn means that some products will have repeated values of ``product complexities''.} 

Now, the values of ECI have been shown to be positively associated with income levels and income growth of countries \citep{HidalgoHausmann2009}. A clear and direct interpretation of the physical meaning of ECI has been lacking, and this has obscured its connection to measures of collective knowhow and economic growth. The reason for this confusion is born out, first, from its flawed interpretation as a \emph{direct} measure of knowhow, and second, from the confusion about its uniqueness.


\newpage
\section*{Acknowledgments}
These notes were written with the intention of serving as teaching material for students and researchers interested in Economic Complexity. They represent just a small portion of a large research agenda whose contributors spread across the globe, and some of the results presented here have widespread origin. However, particular recognition goes to the Growth Lab at the Center for International Development at Harvard University, its director Ricardo Hausmann and research director Frank Neffke, and the conversations with all of CID's researchers, fellows and visitors, where these ideas are discussed, analyzed, debated, and advanced on a daily basis. 

\begin{footnotesize}
\bibliographystyle{apalike}
\bibliography{Ref,Refs20180703}

\begin{thebibliography}{}

\bibitem[Cha, 2007]{cha2007comprehensive}
Cha, S.-H. (2007).
\newblock Comprehensive survey on distance/similarity measures between
  probability density functions.
\newblock {\em City}, 1(2):1.

\bibitem[Coifman and Hirn, 2014]{coifman_diffusion_2014}
Coifman, R.~R. and Hirn, M.~J. (2014).
\newblock Diffusion maps for changing data.
\newblock {\em Applied and Computational Harmonic Analysis}, 36(1):79--107.

\bibitem[Coifman and Lafon, 2006]{coifman_diffusion_2006}
Coifman, R.~R. and Lafon, S. (2006).
\newblock Diffusion maps.
\newblock {\em Applied and Computational Harmonic Analysis}, 21(1):5--30.

\bibitem[Coifman et~al., 2005]{coifman2005geometric}
Coifman, R.~R., Lafon, S., Lee, A.~B., Maggioni, M., Nadler, B., Warner, F.,
  and Zucker, S.~W. (2005).
\newblock Geometric diffusions as a tool for harmonic analysis and structure
  definition of data: Diffusion maps.
\newblock {\em Proceedings of the national academy of sciences},
  102(21):7426--7431.

\bibitem[DeGroot, 1974]{degroot1974reaching}
DeGroot, M.~H. (1974).
\newblock Reaching a consensus.
\newblock {\em Journal of the American Statistical Association},
  69(345):118--121.

\bibitem[Gomez-Lievano et~al., 2016]{gomez2016explaining}
Gomez-Lievano, A., Patterson-Lomba, O., and Hausmann, R. (2016).
\newblock Explaining the prevalence, scaling and variance of urban phenomena.
\newblock {\em Nature Human Behaviour}, 1:0012.

\bibitem[Hausmann et~al., 2011]{hausman2011atlas}
Hausmann, R., Hidalgo, C., Bustos, S., Coscia, M., Chung, S., Jimenez, J.,
  Simoes, A., and Yildirim, M. (2011).
\newblock The atlas of economic complexity.

\bibitem[Hausmann and Hidalgo, 2011]{HausmannHidalgo2011}
Hausmann, R. and Hidalgo, C.~A. (2011).
\newblock The network structure of economic ouput.
\newblock {\em Journal of Economic Growth}, 16:309--342.

\bibitem[Henrich, 2015]{henrich2015secret}
Henrich, J. (2015).
\newblock {\em The secret of our success: how culture is driving human
  evolution, domesticating our species, and making us smarter}.
\newblock Princeton University Press.

\bibitem[Hidalgo et~al., 2018]{hidalgo2018principle}
Hidalgo, C.~A., Balland, P.-A., Boschma, R., Delgado, M., Feldman, M., Frenken,
  K., Glaeser, E., He, C., Kogler, D.~F., Morrison, A., et~al. (2018).
\newblock The principle of relatedness.
\newblock In {\em International Conference on Complex Systems}, pages 451--457.
  Springer.

\bibitem[Hidalgo and Hausmann, 2009]{HidalgoHausmann2009}
Hidalgo, C.~A. and Hausmann, R. (2009).
\newblock The building blocks of economic complexity.
\newblock {\em PNAS}, 106(25):10570--10575.

\bibitem[Hidalgo et~al., 2007]{HidalgoEtAl2007}
Hidalgo, C.~A., Klinger, B., Barabasi, A.-L., and Hausmann, R. (2007).
\newblock {The product space conditions the development of nations}.
\newblock {\em {Science}}, 317(5837):482--487.

\bibitem[Kemp-Benedict, 2014]{kemp2014interpretation}
Kemp-Benedict, E. (2014).
\newblock An interpretation and critique of the method of reflections.
\newblock In {\em Munich Personal RePEc Archive}.

\bibitem[Lafon and Lee, 2006]{lafon_diffusion_2006}
Lafon, S. and Lee, A.~B. (2006).
\newblock Diffusion maps and coarse-graining: A unified framework for
  dimensionality reduction, graph partitioning, and data set parameterization.
\newblock {\em {IEEE} transactions on pattern analysis and machine
  intelligence}, 28(9):1393--1403.

\bibitem[Mealy et~al., 2017]{mealy2017new}
Mealy, P., Farmer, J., and Teytelboym, A. (2017).
\newblock A new interpretation of the economic complexity index.
\newblock {\em {arXiv preprint arXiv:1711.08245}}.

\bibitem[Neffke, 2017]{neffke2017coworker}
Neffke, F. (2017).
\newblock Coworker complementarity.
\newblock In {\em CID Working Papers, no. 79}.

\bibitem[Neffke and Henning, 2013]{Neffke2013SkillRelatedness}
Neffke, F. and Henning, M. (2013).
\newblock Skill relatedness and firm diversification.
\newblock {\em Strategic Management Journal}, 34(3):297--316.

\bibitem[Reshef et~al., 2011]{reshef2011detecting}
Reshef, D.~N., Reshef, Y.~A., Finucane, H.~K., Grossman, S.~R., McVean, G.,
  Turnbaugh, P.~J., Lander, E.~S., Mitzenmacher, M., and Sabeti, P.~C. (2011).
\newblock Detecting novel associations in large data sets.
\newblock {\em science}, 334(6062):1518--1524.

\bibitem[Schaub et~al., 2018]{schaub2018structured}
Schaub, M.~T., Delvenne, J.-C., Lambiotte, R., and Barahona, M. (2018).
\newblock Structured networks and coarse-grained descriptions: a dynamical
  perspective.
\newblock {\em arXiv preprint arXiv:1804.06268}.

\end{thebibliography}
\end{footnotesize}

\end{document}